# Giant Ferroelectric Polarization of CaMn$_7$O$_{12}$ Induced by a Combined Effect of Dzyaloshinskii-Moriya Interaction and Exchange Striction

X. Z. Lu[1], M.-H. Whangbo[2], Shuai Dong[3], X. G. Gong[1], and H. J. Xiang[1,*]

[1] Key Laboratory of Computational Physical Sciences (Ministry of Education), State Key Laboratory of Surface Physics, and Department of Physics, Fudan University, Shanghai 200433, P. R. China

[2] Department of Chemistry, North Carolina State University, Raleigh, North Carolina 27695-8204, USA

[3] Department of Physics, Southeast University, Nanjing 211189, P. R. China

**Abstract**

By extending our general spin-current model to non-centrosymmetric spin dimers and performing density functional calculations, we investigate the causes for the helical magnetic order and the origin of the giant ferroelectric polarization of CaMn$_7$O$_{12}$. The giant ferroelectric polarization is proposed to be caused by the symmetric exchange striction due to the canting of the Mn$^{4+}$ spin arising from its strong Dzyaloshinskii-Moriya (DM) interaction. Our study suggests that CaMn$_7$O$_{12}$ may exhibit a novel magnetoelectric coupling mechanism in which the magnitude of the polarization is governed by the exchange striction, but the direction of the polarization by the chirality of the helical magnetic order.

A crucial issue to solve in the field of spintronics is how to control the magnetism of a solid effectively by electric field. Prospective candidates that can potentially host a strong magnetoelectric (ME) effect are multiferroics in which both magnetic and ferroelectric orders can coexist to host a strong magnetoelectric (ME) effect [1,2]. In particular, those with polarizations driven by a magnetic order are promising because of their intrinsic ME coupling. A direct coupling between magnetism and ferroelectricity was demonstrated in several multiferroics [3-8], but the magnitudes of their polarizations are usually rather small. Very recently, it was reported [9,10] that a mixed-valent manganate $CaMn_7O_{12}$, consisting of one $Mn^{4+}$ and six $Mn^{3+}$ ions per formula unit (FU), exhibits a giant ferroelectric polarization (2870 $\mu C/m^2$) along the c direction at 90 K, below which it adopts a helical magnetic order with propagation vector (0, 1, 0.963). This giant ferroelectric polarization in $CaMn_7O_{12}$ is puzzling: according to the spin current model of Katsura *et al*. [11], the helical magnetic structure cannot induce a nonzero ferroelectric polarization. Recently, we presented a more general model [12] that explains the ferroelectric polarizations induced by a helical magnetic structure. These ferroelectric polarizations arise from spin-orbit coupling (SOC) and hence are very weak in general. Currently, the microscopic origin of the giant ferroelectric polarization in $CaMn_7O_{12}$ is unknown, although a phenomenological ferroaxial coupling mechanism [10,13] has been proposed.

In this Letter, we show on the basis of first principles density functional theory (DFT) calculations that the giant ferroelectric polarization originates mainly from the symmetric exchange striction associated with a particular spin exchange path between $Mn^{4+}$ and $Mn^{3+}$ ions, the Dzyaloshinskii-Moriya (DM) antisymmetric interaction [14] between them is unusually strong for magnetic insulators (i.e., $|D/J| \approx 0.54$ compared with $|D/J| \leq \sim 0.1$ usually expected [14]) [14]), and $CaMn_7O_{12}$ exemplifies a novel ME coupling mechanism.

Above ~440 K $CaMn_7O_{12}$ has a distorted perovskite structure in which, per FU, one $Ca^{2+}$ and three $Mn^{3+}$ (Mn1) ions occupy the A sites with the remaining four $Mn^{3.25+}$ ions at the B sites. On cooling below ~440 K $CaMn_7O_{12}$ undergoes a structural phase transition adopting the space group $R\bar{3}$ [15], in which the four $Mn^{3.25+}$ ions per FU undergo a charge order into one $Mn^{4+}$ (Mn3) and three $Mn^{3+}$ (Mn2) ions without breaking the inversion symmetry. The $Mn1^{3+}$ and $Mn2^{3+}$ ions form chains along the c direction (hereafter //c-chains) [Fig. 1(a)] such that, in every three adjacent //c-chains, the interchain Mn1-Mn2 connections form a spiral chain [Figs. 1(a) and 1(b)]. The spiral chains of $Mn^{3+}$ ions share their //c-chains to form hexagonal tunnels [Fig. 1(c)], which are each occupied by a chain of alternating $Mn^{4+}$ and $Ca^{2+}$ ions so that the $Mn^{4+}$ and $Ca^{2+}$ ions are each surrounded by six //c-chains of $Mn^{3+}$ ions [Fig. 1(d)]. The neutron diffraction measurements [10] show that in the helical magnetic state in the temperature range $T_{N2}$ (48 K) < T < $T_{N1}$ (90 K), the $Mn^{3+}$ spins of each //c-chain are nearly perpendicular to the c-axis and are ferromagnetic (FM) [Fig. 1(e)]. In each spiral chain made up of three //c-chains, the spins of the three FM chains have a compromised arrangement with 120° between the spins of adjacent //c-chains [Fig. 1(f)], showing the presence of spin frustration between them. The $Mn^{4+}$ spins, which are nearly perpendicular to the c-axis, make an angle of ~90° with the $Mn^{3+}$ spins in one set of the three FM chains [dotted triangle in Fig. 1(f)] and ~30° with the $Mn^{3+}$ spins in another set of the three FM chains [dashed triangle in Fig. 1(f)] [hereafter, the (90°, 30°) spin arrangement].

To account for the observed magnetic structure of $CaMn_7O_{12}$ below 90 K, we first evaluate various symmetric spin exchange interactions between the $Mn1^{3+}$, $Mn2^{3+}$ and $Mn3^{4+}$ ions. These ions form the $Mn1O_4$ square planes, the axially compressed $Mn2O_6$ octahedra, and the Mn3O$_6$ octahedra [Figs. 2(a), 2(b) and 2(c)], so that their d-states are split as depicted in Fig. 2(d) according to the computed partial density of states (see Part 3 of [16]). If the distance between the magnetic ions in a spin dimer is restricted to be shorter than 3.7 Å, there are seven different spin exchange paths $J_1 - J_7$ between the Mn1$^{3+}$, Mn2$^{3+}$ and Mn3$^{4+}$ ions (see Fig. S1 of [16]). We evaluate the values of the spin exchanges $J_1 - J_7$ by performing the energy-mapping analysis [17] on the basis of DFT+U calculations (see Part 1 of [16]). The justification for the use of U = 2 and 3 eV in our calculations are given in Part 8 of [16]. Unless mentioned otherwise, results from our calculations with U = 3 eV are presented in the following. The exchange $J_1$ between adjacent Mn1$^{3+}$ and Mn2$^{3+}$ ions in a //c-chain [Fig. 2(a)] is strongly FM ($J_1$ = -5.57 meV, which is an effective spin exchange obtained by setting $|\mathbf{S}_i|$ = 1, namely, $J_{ij}^{eff} = J_{ij}S_iS_j$ for a spin dimer ij). The hybridization between the occupied $t_{2g}$ states of one Mn ion and the empty $e_g$ states of the neighboring Mn ion is stronger for the FM than for the AFM spin arrangement because the energy difference between the occupied and empty d-states is smaller for the FM arrangement, thereby leading to FM $J_1$, which is responsible for the FM arrangement of the Mn$^{3+}$ spins in each //c-chain. The exchange $J_2$ between Mn1$^{3+}$ spins between adjacent //c-chains [Fig. 2(b)] is strongly AFM ($J_2$ = 6.37 meV) due to the Mn-O…O-Mn super-superexchange interactions [18]. Consequently, these inter-chain AFM exchanges cause a strong spin frustration between adjacent FM //c-chains [Fig. 1(g)], and hence the three FM //c-chains in each spiral chain adopt the compromised 120° spin arrangement [Fig. 1(f)].

Our DFT+U+SOC calculations show that the Mn1$^{3+}$ has an easy-axis anisotropy (1.0 meV/Mn) with the easy-axis perpendicular to the Mn1O$_4$ plane [Fig. 2(a)], while the Mn2$^{3+}$ ion has an easy-plane anisotropy (1.5 meV/Mn) with the easy-plane perpendicular to the axially-compressed Mn-O bonds (see Part 4 of [16] for details). As depicted in Fig. 2(a), the easy-axis of the Mn1$^{3+}$ spin and the easy-plane of the Mn2$^{3+}$ spin are much closer to the ab-plane than to the c-axis. The observed orientation of the Mn1$^{3+}$ and Mn2$^{3+}$ spins [Fig. 1(e)] [10], in which all spins are almost perpendicular to the c-axis, is a combined effect of the single-ion anisotropies and the strong FM spin exchange $J_1$ between adjacent Mn1$^{3+}$ and Mn2$^{3+}$ ions.

Let us now consider the preferred orientation of the Mn3$^{4+}$ spin. Each Mn3$^{4+}$ ion is surrounded by six Mn1$^{3+}$ ions and also by six Mn2$^{3+}$ ions [Figs. 1(d), 3(a) and 3(b)]. Thus, each Mn3$^{4+}$ ion has six spin exchanges $J_3$ with the Mn1$^{3+}$ ions and six spin exchanges $J_4$ with the Mn2$^{3+}$ ions [Figs. 1(d) and 2(c)]. Note from Figs. 1(d), 1(e) and 1(f) that the Mn1$^{3+}$ and Mn2$^{3+}$ ions have an identical spin direction if their c-axis heights are the same. In Figs. 3(a) and 3(b) depicting the environment of a Mn$^{4+}$ surrounded by 12 Mn$^{3+}$ ions, the site number 0 refers to the Mn$^{4+}$ ion, the site numbers 1 – 3 to three pairs of Mn1$^{3+}$ and Mn2$^{3+}$ ions with identical spin direction, and the site numbers 4 – 6 to another three pairs of Mn1$^{3+}$ and Mn2$^{3+}$ ions with identical spin direction. For simplicity, we set $|\mathbf{S}_i| = 1$, and define the unit vector $\mathbf{e}_x$ along the direction of $\mathbf{S_1}+\mathbf{S_4}$, and $\mathbf{e}_y$ orthogonal to $\mathbf{e}_x$ in the plane as in Fig. 3(a) so that $\mathbf{e}_z = \mathbf{e}_x \times \mathbf{e}_y$ points toward the reader. Thus, if $\mathbf{S}_1 \times \mathbf{S}_4$ is along $\mathbf{e}_z$, then $\mathbf{S_4}-\mathbf{S_1}$ is along $\mathbf{e}_y$, i.e., $\frac{\mathbf{S_1}-\mathbf{S_4}}{|\mathbf{S_1}-\mathbf{S_4}|} = \mathbf{e}_y \text{sign}[\mathbf{e}_z \cdot (\mathbf{S}_4 \times \mathbf{S}_1)]$. The spins of the two different sets make the angle of 120° between them. Given α as the angle the spin vector $\mathbf{S}_0$ makes with $\mathbf{e}_x$, then the total spin exchange interaction energy $E_{SE}$ of a Mn$^{4+}$ spin with its 12 adjacent Mn$^{3+}$ spins is given by

$$E_{SE} = (J_3 + J_4)\mathbf{S}_0 \cdot (\mathbf{S}_1 + \mathbf{S}_2 + \mathbf{S}_3 + \mathbf{S}_4 + \mathbf{S}_5 + \mathbf{S}_6) = 3(J_3 + J_4)\cos\alpha. \quad (1)$$

Therefore, as long as the sum $(J_3+ J_4)$ is negative (i.e., net FM), which is indeed the case (see Part 8 of [16]), the lowest energy occurs for α = 0°, i.e., for the (60°, 60°) arrangement of the Mn3$^{4+}$ spins. This argument is confirmed by direct DFT+U calculations for the energy $E_{SE}(\alpha)$ of $CaMn_7O_{12}$ with the $Mn1^{3+}$ and $Mn2^{3+}$ spins fixed at the experimentally observed orientations but the spin orientation of the $Mn3^{4+}$ ions varied as a function of the angle $\alpha$. We find the minimum of $E_{SE}(\alpha)$ at $\alpha = 0°$ [Fig. 3(c)], consistent with the above analysis, but in disagreement with the experimental finding that the minimum of $E_{SE}(\alpha)$ occurs at $|\alpha| \approx 30°$ [i.e., for the (90°, 30°) arrangement] [10]. DFT+U+SOC calculations show the energy minimum of $E_{SE}(\alpha)$ -at $\alpha_m = -11°$ [Fig. 3(c)]-in qualitative agreement with experiment indicating the DM interactions to be responsible for the (90°, 30°) spin arrangement of the $Mn3^{4+}$ ions (see below). We point out that $\alpha_m = -28°$ is obtained from DFT+U+SOC calculations with U = 2 eV, in good agreement with experiment (see Part 8 of [16]).

We now examine the ferroelectric polarization of $CaMn_7O_{12}$ by simulating the experimental helical magnetic state with the commensurate helical state **k**= (0, 1, 1) in terms of the hexagonal unit cell. Our DFT+U+SOC calculations show that this helical state has a band gap of 0.45 eV and is more stable than the FM state by 18 meV per FU. The ferroelectric polarization of this helical state with $|\alpha| \approx 30°$ is along the z direction (i.e., c-direction) with $P_z = 4496$ μC/m$^2$ from DFT+U+SOC calculations, but $P_z = 3976$ μC/m$^2$ from DFT+U calculations. Consequently, the giant ferroelectric polarization of $CaMn_7O_{12}$ is caused mainly by exchange striction rather than by SOC.

Given the above finding, it is important to probe which symmetric spin exchange interaction is crucial for the large ferroelectric polarization. Thus we first extend our general spin current model for ferroelelctric polarization [12] to include spin dimers with no centrosymmetric symmetry because the spin dimers of $CaMn_7O_{12}$ are non-centrosymmetric (see Part 5 of [16]).

For a spin dimer containing two spin sites 1 and 2 with no inversion symmetry at the center, the polarization $\mathbf{P_{12}}$ induced by the spin arrangement ($\mathbf{S_1}$, $\mathbf{S_2}$) in the absence of SOC effect can be written as the usual symmetric exchange striction term $\mathbf{P}_{12}(\mathbf{S}_1, \mathbf{S}_2) = \mathbf{P}_{es}(\mathbf{S}_1 \cdot \mathbf{S}_2)$. For the seven exchange paths of the experimental CaMn$_7$O$_{12}$ structure, we evaluate their $\mathbf{P}_{es}$ by performing DFT+U calculations using an energy-mapping method similar to that used to extract the spin exchange parameters [17]. Our calculations show that two exchange paths $J_4$ and $J_5$ have the largest coefficients, namely, $\mathbf{P}_{es}^4 = (-0.024, -0.042, 0.029)$ eÅ and $\mathbf{P}_{es}^5 = (-0.026, -0.048, 0.054)$ eÅ. The remaining spin exchange paths lead to much smaller coefficients. The contribution of $\mathbf{P}_{es}^5$ to the total electric polarization vanishes by symmetry, but $\mathbf{P}_{es}^4$ has a large contribution to the total electric polarization. The large $\mathbf{P}_{es}^4$ arises not only from the small energy gap between the occupied Mn2$^{3+}$ d$_{x2-y2}$↑ state and the empty Mn3$^{4+}$ e$_g$ states in the FM arrangement of the spins in the $J_4$ path but also from the large ∠Mn3-O-Mn2 angle (137.6°), because both reinforce the interaction between the occupied and empty states [Fig. S5(a) and (b) of [16]]. Note from Fig. 1(f) that in the (90°, 30°) spin arrangement the Mn3$^{4+}$ spins give rise to FM-like (~30°) interactions with the Mn2$^{3+}$ spins of the dashed triangle. The difference electron density map between the FM and AFM coupling cases (Fig. S5(d) of [16]) shows the transfer of some electrons from the Mn2 d$_{x2-y2}$ to the Mn3 d$_{z2}$ state.

To understand the role played by SOC [19] on the spin direction of Mn3$^{4+}$, we calculate the DM vectors associated with the seven spin exchange paths $J_1 - J_7$ using our energy-mapping method [17]. The DM vector for the Mn2$^{3+}$ and Mn3$^{4+}$ ions in the exchange path $J_4$ is anomalously large, namely, $|\mathbf{D}_4| = 1.61$ meV and $D_4^z = 1.36$ meV. The latter is about 54% of the symmetric exchange interaction $J_4$ (see Table S1 of [16]). Why the DM interaction $D_4$ is so strong is discussed in Part 7 of [16]. The DM vector for the $Mn2^{3+}$ ions in the exchange path $J_5$ is also relatively large but is not relevant for determining the $Mn3^{4+}$ spin direction. To see if the DM interaction associated with $\mathbf{D}_4$ is indeed responsible for the $Mn3^{4+}$ spin direction, we now write the total spin interaction energy $E_{tot}$ of $Mn3^{4+}$ with its 12 neighboring $Mn^{3+}$ ions as $E_{tot} = E_{SE} + E_{DM}$, where $E_{SE}$ is given by Eq. (1), and the DM interaction energy $E_{DM}(\alpha)$ by

$$E_{DM}(\alpha) = \mathbf{D}_{01} \cdot [\mathbf{S}_0 \times (\mathbf{S}_1 + \mathbf{S}_4)] + \mathbf{D}_{02} \cdot [\mathbf{S}_0 \times (\mathbf{S}_2 + \mathbf{S}_5)] + \mathbf{D}_{03} \cdot [\mathbf{S}_0 \times (\mathbf{S}_3 + \mathbf{S}_6)]$$
$$= 3 D_4^z \sin\alpha. \qquad (2)$$

In deriving the above expression, use was made of the fact that $\mathbf{D}_{01} = \mathbf{D}_{04}$ because the spin dimers 0-1 and 0-4 are related by the inversion symmetry, and $D_{01}^z = D_{02}^z = D_{03}^z$ due to the three-fold rotational symmetry. Therefore,

$$E_{tot}(\alpha) = 3(J_3 + J_4)\cos\alpha + 3 D_4^z \sin\alpha \qquad (3)$$

so the minimum of $E_{tot}(\alpha)$ is obtained when the $Mn3^{4+}$ spin is along the direction given by $\alpha_m = \arctan[D_4^z / (J_3 + J_4)]$. Our calculations show $D_4^z = 1.36\,\mathrm{meV}$, $J_3 = -3.92\,\mathrm{meV}$, $J_4 = -2.96$ meV so that $\alpha_m = -11.2°$, which agrees with the result from the direct first principles calculations [With the $J_3$, $J_4$ and $D_4^z$ parameters determined from DFT+U+SOC calculations with U = 2 eV, we obtain $\alpha_m = -36°$ in closer agreement with $-28°$ from the direct DFT calculations (see Part 8 of [16])]. With the definition of the local coordinate system, $\alpha_m$ does not depend on the chirality and the propagation vector of the helical state.

It should be noted that the spin orientation of $Mn3^{4+}$ is also responsible for the strong ferroelectric polarization arising from the exchange striction. Due to the three-fold rotational symmetry, the total polarization is along z. The polarization per $Mn3^{4+}$ from the exchange striction mechanism can be expressed as

$$\begin{aligned}
P_z &= P_{es,4}^z [\mathbf{S}_0 \cdot (\mathbf{S}_1 - \mathbf{S}_4 + \mathbf{S}_2 - \mathbf{S}_5 + \mathbf{S}_3 - \mathbf{S}_6)] \\
&= 3P_{es,4}^z \mathbf{S}_0 \cdot (\mathbf{S}_1 - \mathbf{S}_4) \\
&= 3P_{es,4}^z |\mathbf{S}_4 - \mathbf{S}_1| \left[ (\cos\alpha)\mathbf{e}_x + (\sin\alpha)\mathbf{e}_y \right] \cdot \mathbf{e}_y \, \text{sign}[\mathbf{e}_z \cdot (\mathbf{S}_4 \times \mathbf{S}_1)] \\
&= 3\sqrt{3} P_{es,4}^z (\sin\alpha) \text{sign}[\mathbf{e}_z \cdot (\mathbf{S}_4 \times \mathbf{S}_1)]
\end{aligned} \quad (4)$$

Thus the magnitude of $P_z$ depends almost linearly on $\alpha$, because $\sin\alpha \approx \alpha$ for small $\alpha$. This finding is consistent with the direct DFT calculations [Fig. 3(c)]. When $\alpha = -30°$, the polarization becomes $-4000$ $\mu C/m^2$. Thus, the large electric polarization originates from the combined effect of the exchange striction and DM interaction.

It is important to note from Eq. (4) that the direction of the polarization depends on the scalar chirality $\sigma \propto \mathbf{e}_z \cdot (\mathbf{S}_4 \times \mathbf{S}_1) \propto \mathbf{r}_{41} \cdot (\mathbf{S}_4 \times \mathbf{S}_1)$. Due to the presence of the DM interaction, two equivalent states with opposite electric polarizations must have the opposite spin chirality. Thus, the polarization reversal, induced by switching the electric field direction, will cause the chirality reversal of the helical spiral. This usually occurs in a multiferroic system such as $LiCu_2O_2$ [20] where the ferroelectricity is due purely to the SOC effect. In the case of $CaMn_7O_{12}$, however, the mechanism of switching the spin chirality is different: The magnitude of the ferroelectric polarization in $CaMn_7O_{12}$ is determined by the exchange striction (thus could be large), but the sign of the polarization by that of the chirality $\sigma$ of the helical magnetic structure due to the strong DM interaction. In the ferroaxial mechanism of Johnson *et al.* [10,13], both the magnitude and the sign of the polarization are determined by the chirality $\sigma$ of the helical magnetic structure.

Work at Fudan was partially supported by NSFC, Pujiang plan, the Special Funds for Major State Basic Research, Foundation for the Author of National Excellent Doctoral Dissertation of China, Research Program of Shanghai municipality and MOE. S.D. was supported by NSFC and 973 Projects of China.

e-mail: hxiang@fudan.edu.cn

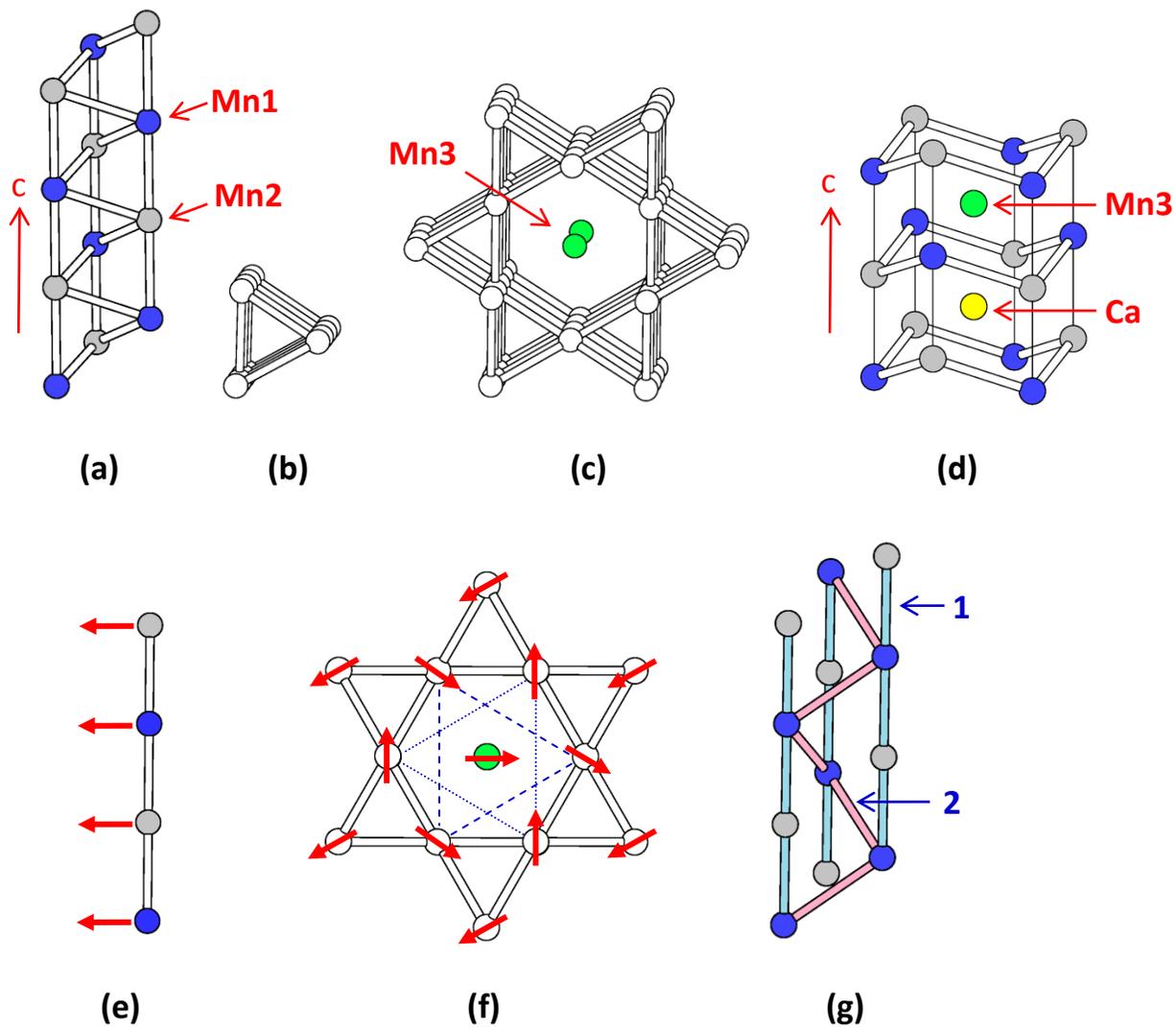

FIG. 1 (color online). (a) Three adjacent //c chains of $Mn^{3+}$ ions, where the blue and grey circles represent the Mn1 and Mn2 atoms, respectively. (b) A spiral chain made up of three adjacent //c-chains of $Mn^{3+}$ ions viewed approximately along the c direction. (c) Three-dimensional arrangement of the $Mn^{3+}$ and $Mn^{4+}$ ions in $CaMn_7O_{12}$ viewed approximately along the c direction. For simplicity, the $Ca^{2+}$ ions are not shown. (d) Arrangements of the $Mn^{3+}$ ions surrounding two adjacent $Mn^{4+}$ and $Ca^{2+}$ ions, where green and yellow circles represent the $Mn3^{4+}$ and $Ca^{2+}$ ions, respectively, and the blue and grey circles the Mn1$^{3+}$ and Mn2$^{3+}$ ions, respectively. (e) Arrangement of the Mn$^{3+}$ spins in a single //c chain. (f) Projection view of the Mn$^{3+}$ and Mn$^{4+}$ spin arrangements in CaMn$_7$O$_{12}$ along the c direction, where each triangle represents three //c-chains forming a spiral chain. The spins of each FM //c-chain are represented by a single spin. (g) Two important spin exchange paths in a spiral chain made up of three adjacent //c-chains, where the numbers 1 and 2 refer to $J_1$ and $J_2$, respectively. $J_1$ is FM, and $J_2$ is AFM.

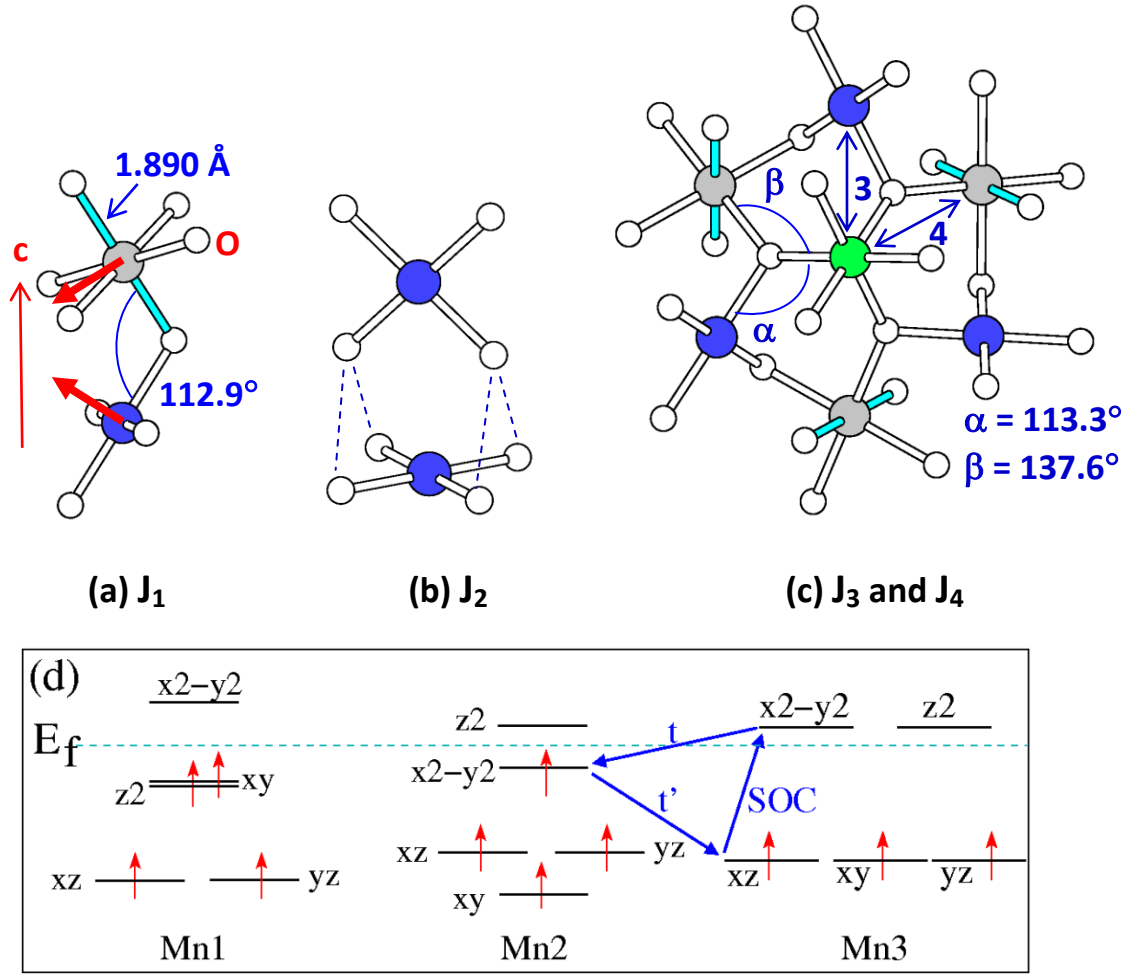

FIG. 2 (color online). (a) The spin exchange path $J_1$ between adjacent $Mn1^{3+}$ (blue circle) and $Mn2^{3+}$ (grey circle) ions in each //c-chain. (b) The spin exchange path $J_2$ between $Mn1^{3+}$ ions that occurs between adjacent //c-chains, where the dotted lines are the O…O contacts shorter than the van der Waals radii sum 3.04 Å (namely, 2.708, 2.727, 2.763 and 2.831 Å). (c) The exchange path $J_3$ between $Mn1^{3+}$ (blue circle) and $Mn3^{4+}$ (green circle), and the exchange path $J_4$ between $Mn2^{3+}$ (grey circle) and $Mn3^{4+}$. (d) The d-state split patterns of the $Mn1O_4$ square plane, the axially-compressed $Mn2O_6$ octahedron, and the $Mn3O_6$ octahedron that best describe the partial density of states [16]. The blue arrows indicate the d-states of $Mn3^{4+}$ and $Mn2^{3+}$ ions involved in the three hopping processes leading to the large DM interaction between adjacent Mn3$^{4+}$ and Mn2$^{3+}$ ions in the spin exchange path $J_4$.

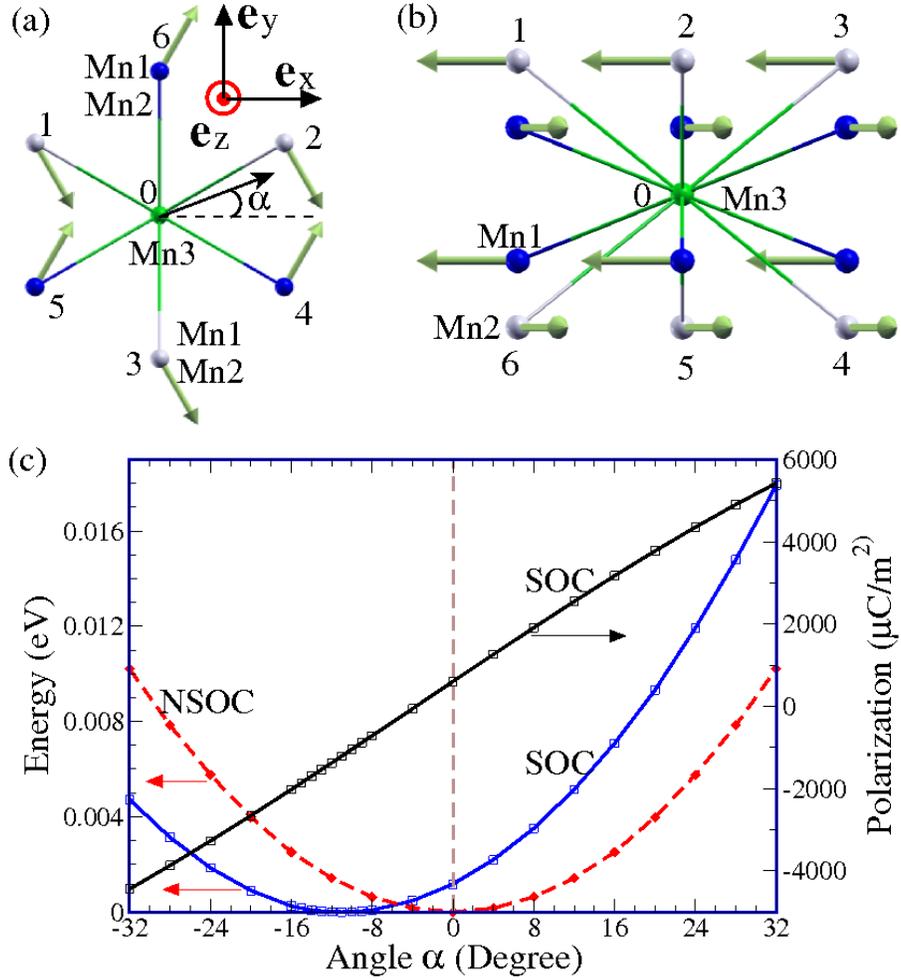

FIG. 3 (color online). (a, b) The top and side views of six $Mn1^{3+}$ and six $Mn2^{3+}$ ions surrounding a $Mn3^{4+}$ ion. The directions of the $Mn1^{3+}$ and $Mn2^{3+}$ spins, observed experimentally, are also shown. (c) The total energy as a function of the $Mn3^{4+}$ spin direction α (see the text for the definition). In the absence of SOC (designated as NSOC), the energy minimum occurs when α= 0°. In the case of SOC, α = −11°.

Supplementary Materials for

**Giant Ferroelectric Polarization of $CaMn_7O_{12}$ Induced by a Combined Effect of Dzyaloshinskii-Moriya Interaction and Exchange Striction**

X. Z. Lu[1], M.-H. Whangbo[2], Shuai Dong[3], X. G. Gong[1], and H. J. Xiang[1*]

[1] Key Laboratory of Computational Physical Sciences (Ministry of Education), State Key Laboratory of Surface Physics, and Department of Physics, Fudan University, Shanghai 200433, P. R. China

[2] Department of Chemistry, North Carolina State University, Raleigh, North Carolina 27695-8204, USA

[3] Department of Physics, Southeast University, Nanjing 211189, P. R. China

# 1. Details of the density functional calculations

Total energy calculations are based on the DFT plus the on-site repulsion (U) method [1] within the generalized gradient approximation [2] (DFT+U) on the basis of the projector augmented wave method [3] encoded in the Vienna ab initio simulation package [4]. The plane-wave cutoff energy is set to 400 eV. SOC is included in the calculations unless noted otherwise. We mainly discuss the results obtained with the on-site repulsion U = 3 eV and the exchange parameter J = 1 eV on Mn. When using a large U such as 4 eV, FM is more stable than the experimental helical spiral state (see Part 6 of Supplementary Materials). For the calculation of electric polarization, the Berry phase method [5] was used. For the reference state for computing the polarization, we use the antiferromagnetic (AFM) state in which all Mn1 spins are up, all Mn2 spins are down, and all Mn3 spins are up. This AFM state preserves the inversion symmetry of the system.

In this study, we mainly discuss the results obtained by using the experimental crystal structure of $CaMn_7O_{12}$. We relax the atomic positions of $CaMn_7O_{12}$ using the experimental spin structure, to find $P_z$ = 3182 $\mu C/m^2$ for the relaxed structure. The latter is close to the value calculated for the experimental structure. Thus, our main conclusions based on the experimental structure should remain valid.

**2. Values of the symmetric exchanges, the DM antisymmetric exchanges, and the exchange striction polarization coefficients calculated for seven pairs of spin sites**

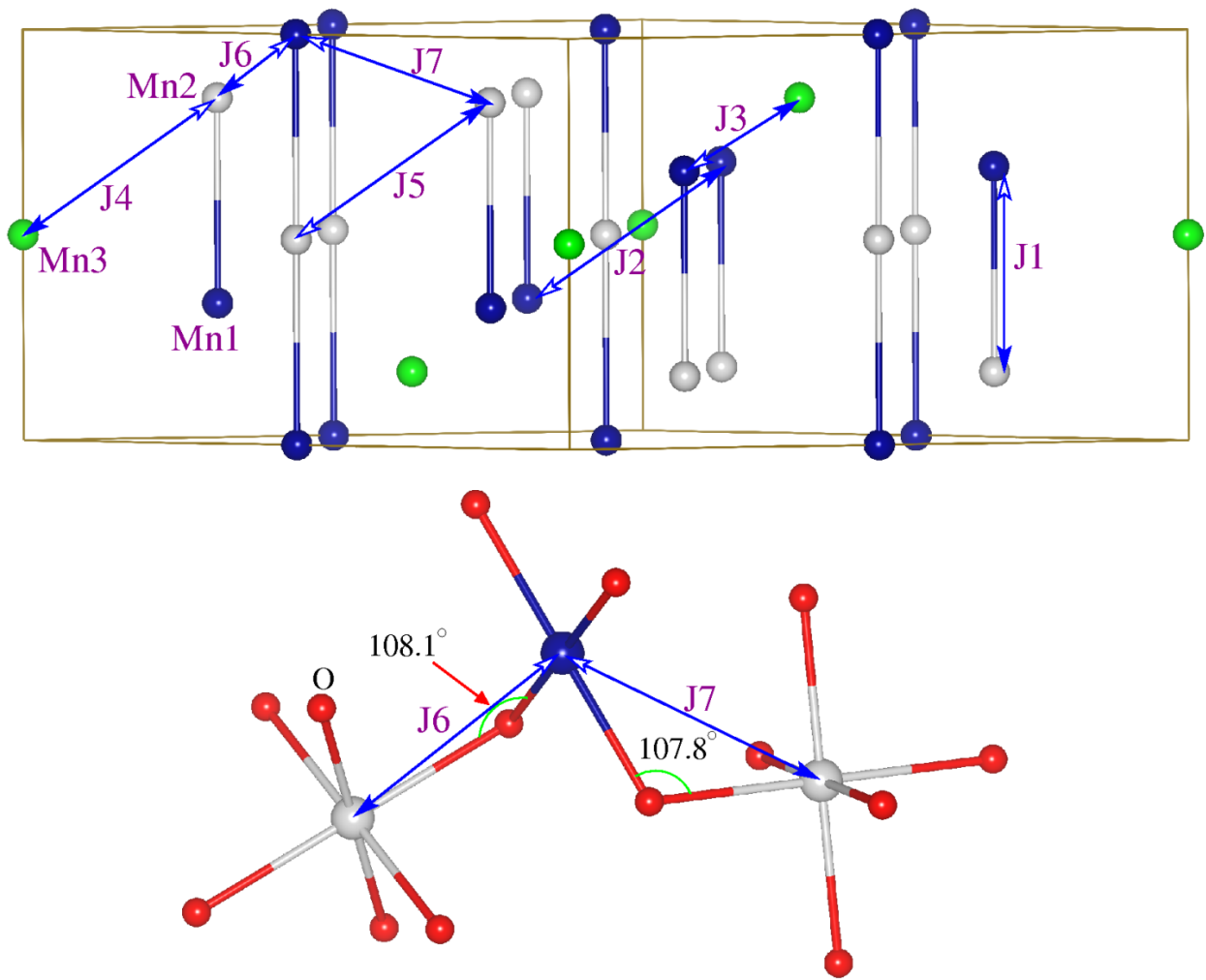

Figure S1 (color online). The seven Mn-Mn pairs with distance shorter than 3.7 Å in the room temperature crystal structure of $CaMn_7O_{12}$ (hexagonal cell). For clarity, we only show the Mn ions in the upper panel (The Mn1-Mn2 1D chains are shown). The detailed structure of the Mn1-Mn2 pairs (J6 and J7) with the same distance ( Å) is shown in the lower panel. In depicting the DM vector $\mathbf{D}_{ij}$, we use the convention in which the empty arrow points toward the site i, while the solid arrow points toward the site j.

Table S1. Calculated symmetric exchange interaction parameters (J), antisymmetric DM interaction parameters (**D**), and symmetric exchange striction coefficients (**P**$_{es}$) for the electric polarization obtained from DFT calculations with U = 3 eV.

| Path | Pair | Distance (Å) | J (meV) | **D** (meV) | **P**$_{es}$ (e.Å) |
|---|---|---|---|---|---|
| 1 | Mn1-Mn2 | 3.1715 | -5.565 | (0.096, -0.115, -0.135) | (0.024, 0.000, 0.008) |
| 2 | Mn1-Mn1 | 3.6817 | 6.368 | (0.001, 0.023, 0.015) | (-0.002, -0.001, 0.005) |
| 3 | Mn1-Mn3 | 3.1941 | -3.916 | (0.142, 0.058, 0.044) | (0.012, 0.009, -0.005) |
| 4 | Mn2-Mn3 | 3.6817 | -2.961 | (-0.064, 0.862, 1.361) | (-0.024, -0.042, 0.029) |
| 5 | Mn2-Mn2 | 3.6817 | -2.534 | (1.243, -0.780, -0.410) | (-0.026, -0.048, 0.054) |
| 6 | Mn1-Mn2 | 3.1941 | -2.290 | (0.121, 0.205, 0.217) | (0.010, 0.020, -0.011) |
| 7 | Mn1-Mn2 | 3.1941 | 2.412 | (-0.022, -0.209, 0.035) | (0.000, -0.016, 0.017) |

## 3. DOS plots of CaMn$_7$O$_{12}$ from first-principles calculations

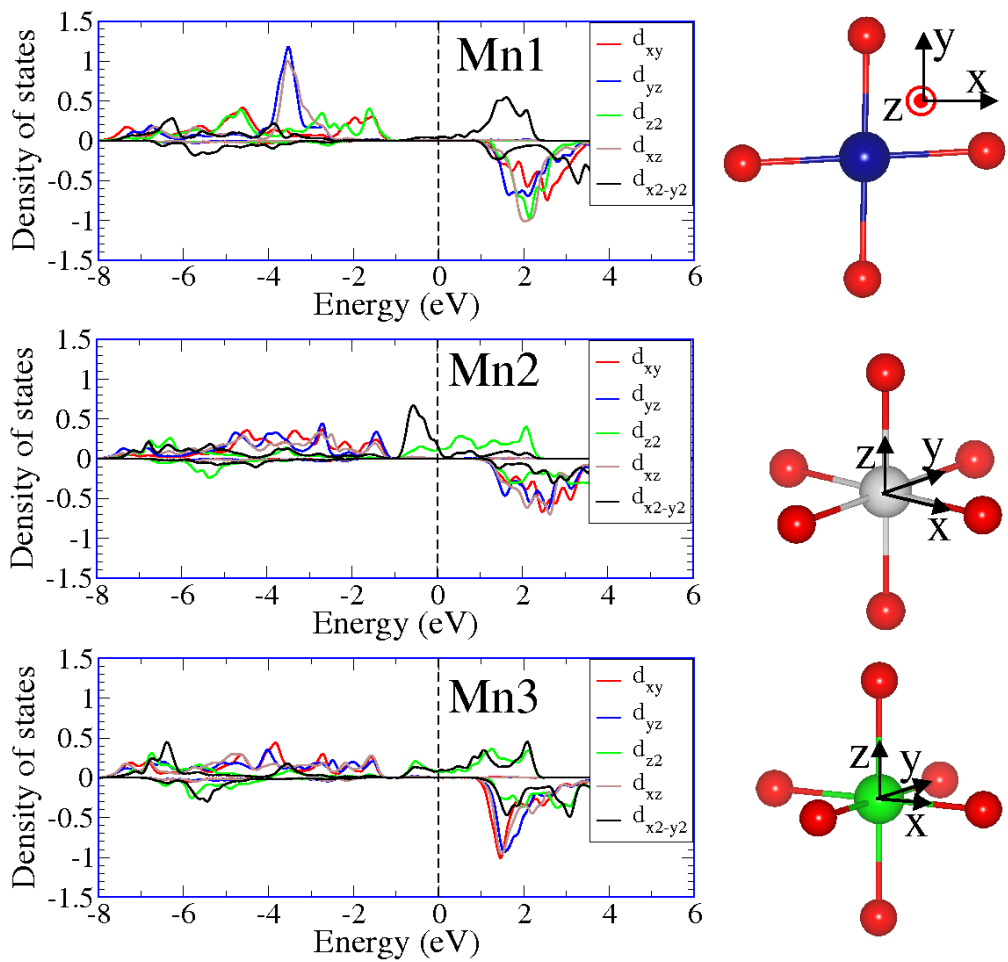

Figure S2 (color online). The PDOS plots calculated for the d-states of Mn1, Mn2, and Mn3 atoms. The FM state is used in the DFT+U calculation with $U = 3$ eV and $J = 1$ eV. The local coordinate systems used for the Mn1, Mn2, and Mn3 atoms are shown in the right panel.

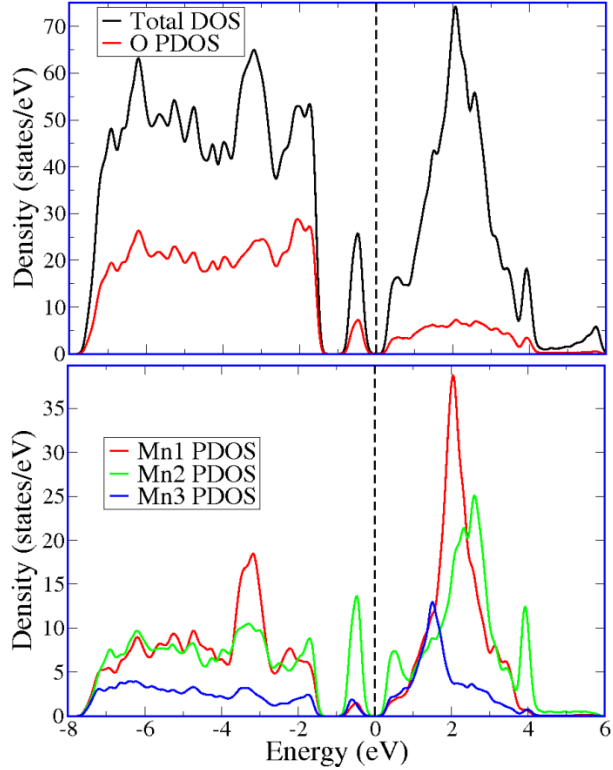

Figure S3 (color online). The partial DOS for Mn and O ions from the DFT+U+SOC calculation with U = 3 eV and J = 1 eV using the experimental helical spin state. The total DOS is also shown.

## 4. Single-ion anisotropy of Mn ions

According to the relativistic quantum mechanics, the one-electron SOC Hamiltonian $H_{SO} = \lambda \mathbf{L} \cdot \mathbf{S}$, where $\lambda$ is the SOC constant and the $\mathbf{L} \cdot \mathbf{S}$ term can be written as [6]:

$$\mathbf{L} \cdot \mathbf{S} = S_n \left( L_z \cos\theta + \frac{1}{2} L_+ e^{-i\phi} \sin\theta + \frac{1}{2} L_- e^{i\phi} \sin\theta \right)$$

$$+ \frac{1}{2} S_+ \left( -L_z \sin\theta - L_+ e^{-i\phi} \sin^2\frac{\theta}{2} + L_- e^{i\phi} \cos^2\frac{\theta}{2} \right)$$

$$+ \frac{1}{2} S_- \left( -L_z \sin\theta + L_+ e^{-i\phi} \cos^2\frac{\theta}{2} - L_- e^{i\phi} \sin^2\frac{\theta}{2} \right),$$

where θ and φ are the zenith and azimuth angles of the magnetization direction **n**. Usually, the spin conserving term $S_n \left( L_z \cos\theta + \frac{1}{2} L_+ e^{-i\phi} \sin\theta + \frac{1}{2} L_- e^{i\phi} \sin\theta \right)$ is the dominant term due to the smaller energy differences in the SOC-induced mixing between occupied and unoccupied states. In the case of Mn1, |xy> can mix with |x2-y2> through the $L_z \cos\theta$ term. When the spin is along the local z axis (i.e., $\theta = 0$), it leads to the lowest energy. The Mn2 case is different: The term, $\frac{1}{2} L_+ e^{-i\phi} \sin\theta + \frac{1}{2} L_- e^{i\phi} \sin\theta$, leads to the mixing between |xz>, |yz>, and |z2>. The energy lowering is the largest when $\theta = 90$, i.e., easy-plane anisotropy. The $Mn3^{4+}$ ion has no significant single-ion anisotropy because it has six equal Mn-O bonds, but has a small orbital moment (~$0.02\mu_B$).

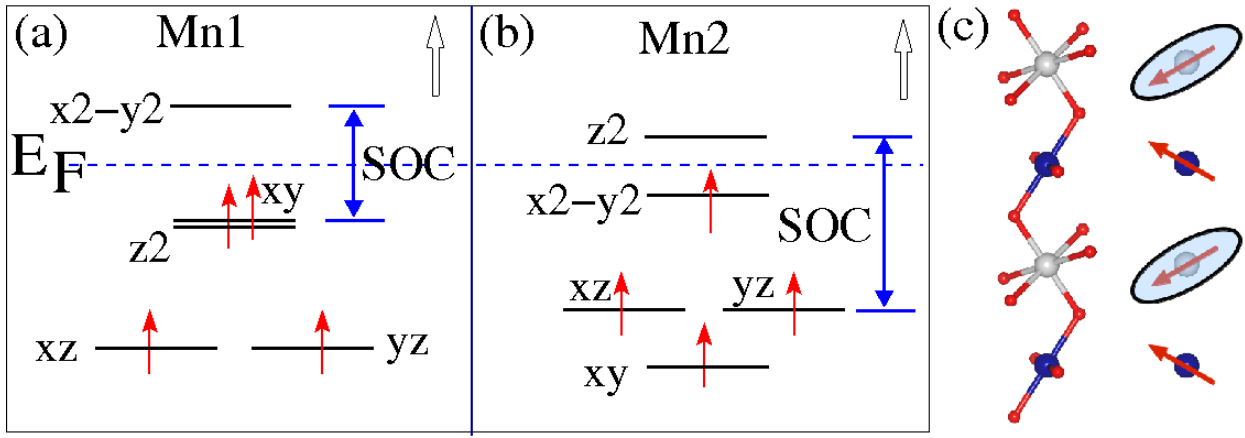

Figure S4 (color online). (a) The mechanism for the easy-axis behavior of Mn1. (b) The mechanism for the easy-plane behavior of Mn2. (c) The arrangement of the easy axis of Mn1 and the easy-plane of Mn2 in a //c-chain of the Mn1 and Mn2 atoms. Both the easy axis and the easy-plane are closer to the ab-plane than to the c-axis. The single-ion anisotropy and the intrachain FM interaction between Mn1 and Mn2 lead to the FM chain with in-plane spins.

**5. Theory of spin-order induced electric polarization including the non-centrosymmetric dimers**

We previously developed a theory of spin-order induced electric polarization for the case of a centrosymmetric dimer [7]. Now, we extend this model [7] to describe the non-centrosymmetric dimer case. Let us first consider a spin dimer without inversion symmetry at the center. Without loss of generality, the propagation vector from spin 1 to spin 2 will be taken along the x-axis. In general, the electric polarization **P** is a function of the directions of spin 1 and spin 2 (with unit vectors **S**$_1$ and **S**$_2$, respectively), namely, **P** = **P**($S_{1x}$, $S_{1y}$, $S_{1z}$, $S_{2x}$, $S_{2y}$, $S_{2z}$). In principle, therefore, **P** can be expanded as a Taylor series of $S_{i\alpha}$ (i = 1, 2; α = x, y, z). The time-reversal symmetry requires that inverting both spin directions leave the electric polarization unchanged. Thus, the odd terms of the Taylor expansion should vanish. If the fourth and higher order terms are neglected, **P** is written as

$$\mathbf{P} = \mathbf{P}_1(\mathbf{S}_1) + \mathbf{P}_2(\mathbf{S}_2) + \mathbf{P}_{12}(\mathbf{S}_1, \mathbf{S}_2), \tag{1}$$

where **P**$_i$(**S**$_i$) (i = 1, 2) is the intra-site polarization and the inter-site polarization **P**$_{12}$(**S**$_1$, **S**$_2$) are given by

$$\mathbf{P}_{12}(\mathbf{S}_1, \mathbf{S}_2) = \sum_{\alpha\beta} \mathbf{P}_{12}^{\alpha\beta} S_{1\alpha} S_{2\beta}. \tag{2}$$

Because the intra-site term is found to be small [7], we hereafter only consider the inter-site term, which can be written in the following form:

$$\begin{aligned}
\mathbf{P}_{12}(\mathbf{S}_1, \mathbf{S}_2) &= \sum_{\alpha\beta} \mathbf{P}_{12}^{\alpha\beta} S_{1\alpha} S_{2\beta} \\
&= (S_{1x}, S_{1y}, S_{1z}) \begin{pmatrix} \mathbf{P}_{12}^{xx} & \mathbf{P}_{12}^{xy} & \mathbf{P}_{12}^{xz} \\ \mathbf{P}_{12}^{yx} & \mathbf{P}_{12}^{yy} & \mathbf{P}_{12}^{yz} \\ \mathbf{P}_{12}^{zx} & \mathbf{P}_{12}^{zy} & \mathbf{P}_{12}^{zz} \end{pmatrix} \begin{pmatrix} S_{2x} \\ S_{2y} \\ S_{2z} \end{pmatrix} \\
&= \mathbf{S}_1^T \mathbf{P}_{int} \mathbf{S}_2
\end{aligned}$$

As in the case of spin exchange interactions [6], **P**$_{int}$ can be written as the sum of an isotropic symmetric diagonal matrix **P**$_J$, an antisymmetric matrix **P**$_D$, and an anisotropic symmetric matrix **P**$_\Gamma$. It should be noted that here all the matrix elements are vectors.

When the SOC effect is absent, any global rotation of all the spins should not change the polarization of the system. In this case, the formula can be drastically simplified. Consider the case with **S**$_1$ = (1,0,0) and **S**$_2$ = (0,1,0) and the case with **S**$_1$ = (1,0,0) and **S**$_2$ = (0,-1,0). The second spin state can be obtained by performing a spin rotation of the spins along the x axis by 180°. Therefore, $\mathbf{P}_{12}^{xy}$ should be zero because these two states have the same polarization. One can also easily see that $\mathbf{P}_{12}^{xx} = \mathbf{P}_{12}^{yy} = \mathbf{P}_{12}^{zz}$. Therefore, we have the usual symmetric exchange striction term $\mathbf{P}_{12}(\mathbf{S}_1, \mathbf{S}_2) = \mathbf{P}_{es}(\mathbf{S}_1 \cdot \mathbf{S}_2)$, where $\mathbf{P}_{es} = \mathbf{P}_J^{11} = \mathbf{P}_J^{22} = \mathbf{P}_J^{33}$. It should be noted that the symmetric exchange striction term vanishes in the centrosymmetric dimer case.

We can extract $\mathbf{P}_{es}$ by using a method similar to the one used to extract the spin exchange parameters proposed in our previous work [8]. Without loss of generality, let us consider the exchange path between site 1 and site 2. We can calculate the electric polarizations using the Berry phase method [5] for the following four states: (I) up-spins for both site 1 and site 2 (↑,↑), (II) up-spin for site 1 and down-spin for site 2 (↑,↓), (III) down-spin for site 1 and up-spin for site 2 (↓,↑), (IV) down-spins for both site 1 and site 2 (↓,↓). In these four spin states, the spin orientations for the spin sites other than 1 and 2 are the same but can be arbitrary. Then the coefficient can be computed as $\mathbf{P}_{es}^{12} = (\mathbf{P}^I + \mathbf{P}^{IV} - \mathbf{P}^{II} - \mathbf{P}^{III})/4$.

## 6. Origin of the large polarization coefficient $\mathbf{P}_{es}^4$

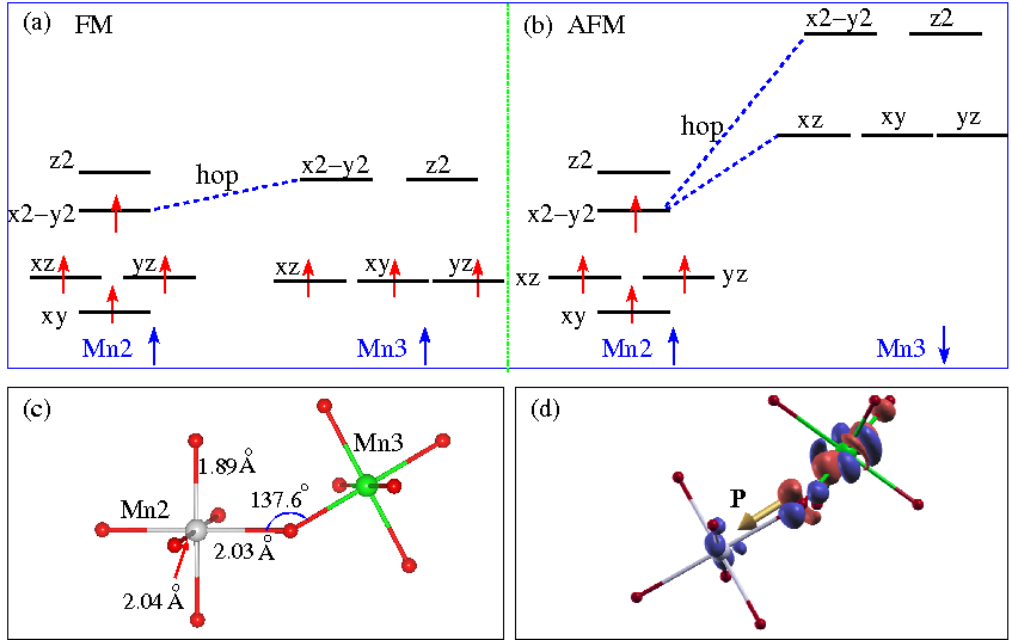

Figure S5 (color online). The interactions between the d-states of the $Mn2^{3+}$ and $Mn3^{4+}$ ions through the Mn2-O-Mn3 linkage in the spin exchange path $J_4$ in the cases when the spins of the two sites have (a) FM and (b) AFM arrangements. (c) The spin dimer structure associated with the exchange path $J_4$ between $Mn2^{3+}$ and $Mn3^{4+}$. (d) The difference electron density plot $\Delta\rho = \rho(\uparrow,\uparrow)+\rho(\downarrow,\downarrow)-\rho(\uparrow,\downarrow)-\rho(\downarrow,\uparrow)$ (see Part 5), where the red color means $\Delta\rho > 0$.

## 7. Reason why the DM interaction $D_4$ is strong

Finally, let us consider why the DM interaction $\mathbf{D}_4$ is so strong. By switching off the SOC

effect in the DFT+SOC+U calculations on the Mn2$^{3+}$ and Mn3$^{4+}$ sites separately, we find that $\mathbf{D}'_4$ = (-0.036, 0.254, 0.473) meV if SOC acts only on Mn2$^{3+}$, $\mathbf{D}''_4$ = (-0.042, 0.607, 0.867) meV if SOC acts only on Mn3$^{4+}$, and $\mathbf{D}_4 \approx \mathbf{D}'_4 + \mathbf{D}''_4$. As shown by Moriya [14], the DM interaction between two spin sites is given by the sum of various three hopping processes between them. Our analysis for $\mathbf{D}_4$ indicates that the processes shown in Fig. 2(d) are the dominant ones; at the Mn3$^{4+}$ site an electron hops from the occupied xz/yz state to the unoccupied $x^2$-$y^2$ state by SOC, and then hops into the $x^2$-$y^2$ state of the neighboring Mn2$^{3+}$ site (with hopping integral $t$) because the electron present in the latter hops into the xz/yz state of the Mn3$^{4+}$ (with hopping integral $t'$) vacated by the SOC-induced hopping. A strong DM interaction results because each of the three steps involved is strong. Note that the hopping integrals $t$ and $t'$ are both large because the ∠Mn3-O-Mn2 angle (137.6°) is close to 135°. Thus, as in the case of the large $\mathbf{P}^4_{es}$, the large $\mathbf{D}_4$ is caused by the small energy gap between the occupied Mn2$^{3+}$ $d_{x2-y2}\uparrow$ state and the unoccupied Mn3$^{4+}$ $e_g$ states in the FM state and also by the large ∠Mn3-O-Mn2 angle.

## 8. Results from the calculations with other U values

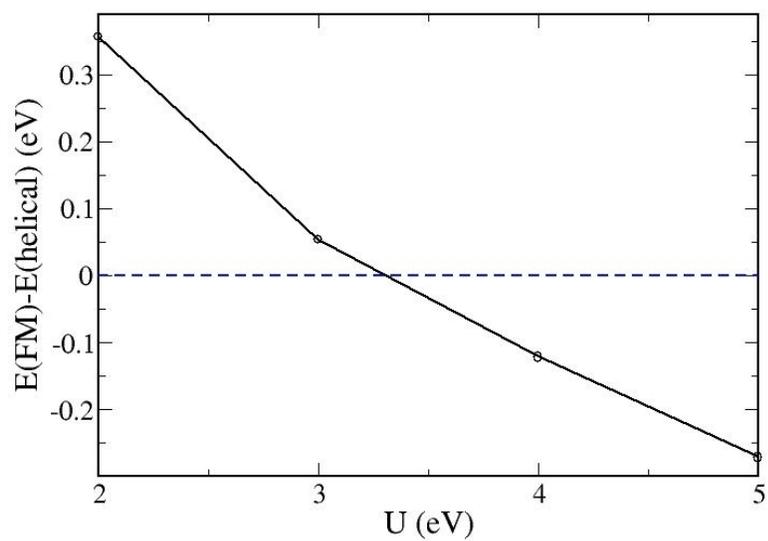

Figure S6 (color online). The energy difference between the FM state and the experimentally observed helical state as a function of the U value from the DFT+U+SOC calculations. When U is smaller than 3.3 eV, the helical state is predicted to be more stable.

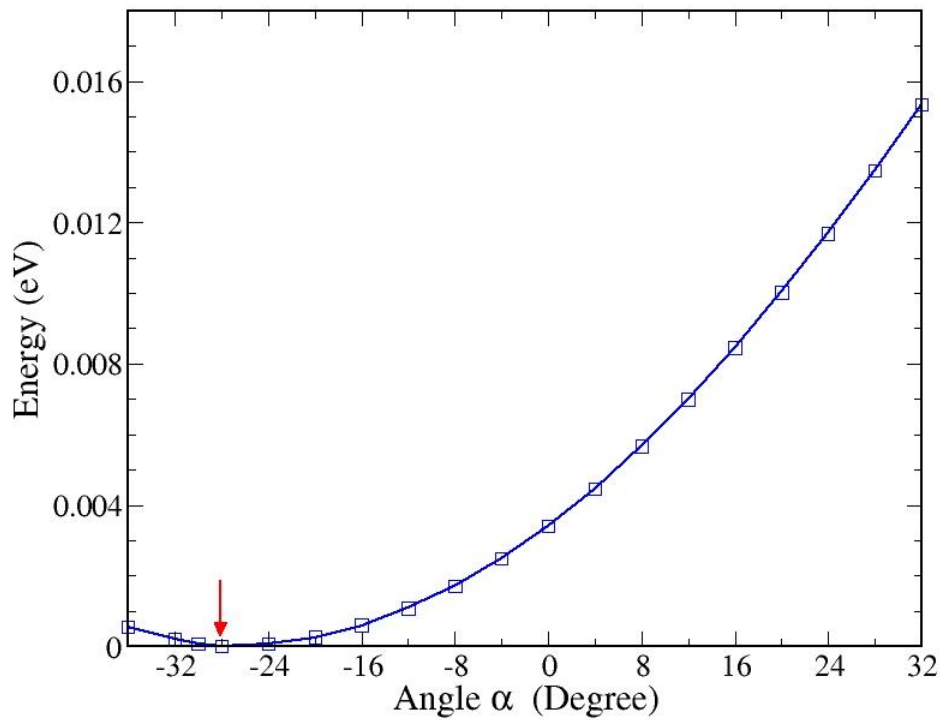

Figure S7 (color online). The total energy as a function of the $Mn3^{4+}$ spin orientation angle $\alpha$ (see the text for the definition) from the DFT+U+SOC calculations with U = 2 eV. The energy minimum occurs at $\alpha = -28°$, which is very close to the experimental value.

In the case of U = 2 eV, our calculations show that $J_3 = -3.117$ meV, $J_4 = 0.846$ meV, $\mathbf{D}_4 =$ (-0.139, 1.064, 1.665) meV. Therefore $J_3+J_4$ is net FM (i.e., negative). The optimal $\alpha_m$ of the $Mn3^{4+}$ spin is given by: $\alpha_m = \arctan[D_4^z/(J_3+J_4)] = -36.2°$, which is close to the result (-28°) from the direct DFT calculations.